\documentclass{article}

\usepackage{PRIMEarxiv}

\usepackage[utf8]{inputenc} 
\usepackage[T1]{fontenc}    
\usepackage{hyperref}       
\usepackage{url}            
\usepackage{booktabs}       
\usepackage{amsfonts}       
\usepackage{nicefrac}       
\usepackage{microtype}      
\usepackage{lipsum}
\usepackage{fancyhdr}       
\usepackage{graphicx}       
\graphicspath{{media/}}     
\usepackage{amsmath}
\usepackage{booktabs} 
\usepackage{multirow}
\usepackage{natbib}
\usepackage[version=4]{mhchem}

\title{Radiative Association of Ag and H: Formation of AgH from Ab Initio Calculations
\thanks{\textit{\underline{Citation}}: 
\textbf{Authors. Title. Pages.... DOI:000000/11111.}} 
}

\author{
  Lin Jiang, Yu Wang, Xuanbing Qiu, Yali Tian, Guqing Guo, Chuanliang Li \\
  School of Applied Science \\
  Taiyuan University of Science and Technology \\
  Taiyuan\\
  \texttt{\{Yu Wang\}wangy@tyust.edu.cn} \\
   \And
   Yukun Yang \\
  School of Physics \\
  Henan Normal University \\
  Xinxiang \\
  \texttt{yangyukun@htu.edu.cn} \\
     \And
   Ling Liu, Yong Wu \\
  Key Laboratory of Computational Physics \\
  Institute of Applied Physics and Computational Mathematics \\
  Beijing \\
  \texttt{\{Yong Wu\}wu\_yong@iapcm.ac.cn} \\
}

\begin{document}
\maketitle

\begin{abstract}
Radiative association processes leading to the formation of AgH in cold astrophysical environments are investigated for the first time using full quantum scattering theory. High accuracy potential energy curves and transition dipole moments for the low-lying electronic states (X$^1\Sigma^+$, A$^1\Sigma^+$, $1^1\Pi$, $3^1\Sigma^+$, $2^1\Pi$) are computed employing the internally contracted multireference configuration interaction method with Davidson correction. Vibrationally and rotationally resolved radiative association cross sections are calculated for transitions from these initial states to the ground X$^1\Sigma^+$ state. Prominent shape resonances arising from quasi-bound rovibrational levels behind centrifugal barriers are identified, with the $2^1\Pi \to$ X$^1\Sigma^+$ channel exhibiting the strongest contribution at low collision energies. Stimulated radiative association under blackbody radiation fields (up to $T = 20\,000$ K) produces modest enhancements, predominantly in the ground-state channel. Thermal rate coefficients computed over 10$^{-1}$--$10^4$~K reveal a general decreasing trend with temperature for all channels. The results provide essential kinetic data for astrochemical models of transition-metal hydride formation in low-temperature interstellar and circumstellar environments.
\end{abstract}


\section{Introduction}

In astrophysical environments, the formation of molecules is a cornerstone of cosmic chemical evolution. Particularly in low-density regions where three-body collisions are statistically improbable, radiative association (RA)—a process in which two colliding atoms emit a photon to form a bound molecule—serves as the dominant pathway for the genesis of diatomic molecules. For instance, RA plays a crucial role in the synthesis of CH$^+$ in interstellar clouds, enabling early molecular chemistry in the universe \cite{Abgrall1976QuantumEI}, and in forming dust nucleation seeds like MgS in carbon-rich asymptotic giant branch (AGB) stars \cite{Wei2025}. Similarly, RA contributes to AlO production in stellar atmospheres, influencing dust grain formation \cite{Di2025}.

Driven by its astrophysical significance, the theoretical framework of RA has evolved substantially. Seminal early works laid the foundation:  Abgrall et al. (1976) illustrated resonance enhancements in CH$^+$ formation driven by rotational quantization \cite{Abgrall1976QuantumEI}; Herbst (1980) introduced statistical models affirming RA's efficacy in low-temperature settings \cite{Herbst1980ApJ}; and Bates (1983) synthesized quantum and statistical approaches, offering simplified cross-section formulas \cite{Bates1983ApJ}. More recent developments have embraced full quantum mechanical computations: Liu et al. (2010) examined competing radiative processes \cite{liu2010radiative}; Stoecklin et al. (2013) refined methodologies for atomic-diatomic systems \cite{stoecklin2013new}; Babb and colleagues (2017, 2019) computed RA for C+H$^+$ and C+C$^+$ systems, uncovering essential channels and database-ready fitting forms \cite{babb2017radiative,babb2019radiative}; Šimsová-Zámečníková et al. (2022) underscored non-adiabatic effects in NaCl \cite{vsimsova2022formation}; and Chen et al. (2024) supplied data for AlCl and PCl$^+$ \cite{chen2024radiative}. However, despite this progress, detailed quantum mechanical studies on transition-metal systems remain conspicuously scarce, with Silver Hydride (AgH) being a notable omission.

 In the specific case of AgH, existing literature has predominantly focused on its spectroscopic characterization and electronic structure. Learner (1962) furnished spectral constants \cite{learner1962influence}; Le Roy et al. (2005) captured high-resolution spectra and potential curves for AgH isotopes \cite{le2005direct}; Li et al. (2018) pinpointed quasi-closed transitions suitable for laser cooling; Alrebdi et al. (2021) evaluated dipole moments and vibrational levels \cite{alrebdi2021ab}; Zhou et al. (2023) analyzed spin-orbit coupling influences \cite{zhou2023theoretical}; and Mohammadian et al. (2024) reclassified excited states while computing Einstein coefficients \cite{mohammadian2024spin}. Nevertheless, RA processes in AgH have yet to be explored. This oversight is noteworthy, given that silver permeates the interstellar medium via supernova ejecta and AGB star outflows \cite{Jenkins2009}, and comparable metal hydrides such as AlH have been observed in stellar atmospheres \cite{Roederer2014}. Thus, examining RA for AgH offers valuable perspectives on the genesis of transition-metal hydrides in cold astrophysical settings.

\section{Method of calculation} \label{S2}

\subsection{Electronic-structure methodology}\label{S2.1}

The electronic structure calculations for AgH were carried out with the \texttt{MOLPRO 2012} package \cite{werner1988efficient,werner2012molpro}. For the Ag atom, the \texttt{aug-cc-pVTZ-PP} basis set was employed together with the Stuttgart ECP28MDF small-core relativistic effective core potential \cite{peterson2003systematically}, which replaces the 28 inner-shell electrons ($\mathrm{1s^22s^22p^63s^23p^63d^{10}}$) and incorporates scalar relativistic effects. The H atom was described using the all-electron \texttt{aug-cc-pVTZ} basis set. This computational setup is consistent with recent high-level studies on AgH and related spectroscopic properties \cite{zhou2023theoretical,mohammadian2024spin}.

All calculations were performed in the $\mathrm{C}_{2v}$ point group, the largest Abelian subgroup of $\mathrm{C}_{\infty v}$, with the symmetry correspondence $\Sigma^+ = A_1$, $\Pi = B_1 + B_2$, $\Delta = A_1 + A_2$, and $\Sigma^- = A_2$. 
The molecular orbitals were first generated at the Hartree--Fock level \cite{werner1980quadratically}, followed by state-averaged CASSCF calculations. The state averaging included five states of $A_1$ symmetry, two states of $B_1$, two states of $B_2$, and one state of $A_2$. The active space comprised 12 valence electrons distributed in 10 active orbitals, corresponding to the Ag 4d, 5s, 5p, 6s and H 1s, 2s, 2p shells. The Ag 4s4p semicore orbitals were kept doubly occupied as closed orbitals, while the deeper core electrons were represented by the effective core potential. On top of the SA-CASSCF reference, internally contracted MRCI calculations with Davidson correction (+Q) were carried out to recover dynamic correlation.

The PECs were computed over the internuclear range 0.5--10~\AA. Transition dipole moments from the excited singlet states to X$^1\Sigma^+$ and the permanent dipole moment of X$^1\Sigma^+$ were evaluated over the same range. For the scattering calculations, the \emph{ab initio} points were interpolated by cubic splines in the intermediate region, extrapolated at short range with an exponential form,
\begin{equation}
V_S(R) = a e^{-b R} + V_{\min},
\end{equation}
where $a$ and $b$ are fitting parameters determined from the slope between the first two ab initio data points near the minimum, ensuring a physically realistic Born-Mayer-type repulsion. For $\mathrm{R} > 10$ \AA{} (long-range), the tail of the PECs was fitted using
\begin{equation}
V_L(R) = \frac{A}{R^6} + \frac{B}{R^8} + \frac{C}{R^{10}},
\label{fitPEC}
\end{equation}
with coefficients $A$, $B$, $C$ obtained by fitting the ab initio data in the long-range region. Similarly, TDMs were fitted as
\begin{equation}
\text{TDM}_L(R) = \frac{D}{R^6} + \frac{E}{R^8} + \frac{F}{R^{10}} + x,
\label{fitDM}
\end{equation}
with coefficients $D$, $E$, $F$ obtained by fitting the ab initio data in the long-range region after shifting the data to ensure asymptotic decay to zero at large distances. Here, $x$ represents the TDM value at the maximum computed $R$, relative to zero (asymptotic limit). This form ensures smooth transitions and physically correct van der Waals behavior in neutral systems, where the leading $C_6$ term dominates, providing smoother extrapolation than forms with additional $C_5$ terms (as in some literature for charged or polar systems, e.g. \cite{Di2025}).

\subsection{Radiative association cross sections and rate coefficients}\label{S2.2}

Radiative association (RA) occurs when two colliding atoms form a quasi-bound complex that stabilizes by emitting a photon. The cross section for spontaneous transitions from an initial electronic state $\Lambda$ to a final state $\Lambda'$ is computed quantum mechanically \cite{zygelman1990radiative}:
\begin{equation}\label{eq1}
\sigma_{\Lambda \to \Lambda'}^{\rm sp}(E) = \frac{64\pi^5}{3 \hbar^3 c^3} \frac{p_\Lambda}{k^2} \sum_{J v' J'} \nu_{E,v',J'}^3 S_{JJ'} |M_{EJ,v'J'}|^2,
\end{equation}
where $E$ is the collision energy, $k$ is the wave number, $\nu_{E,v',J'}$ is the transition frequency with $E_{v'J'}$ the bound state energy, $S_{JJ'}$ is the H\"{o}nl-London factor (listed in Table~\ref{tab:HL}; \citealp{gianturco1996radiative}), and $p_\Lambda$ is the statistical weight:
\begin{equation}\label{eq2}
p_\Lambda = \frac{(2S' + 1)(2 - \delta_{0,\Lambda})}{(2L_A + 1)(2S_A + 1)(2L_B + 1)(2S_B + 1)},
\end{equation}
where $\delta_{0,\Lambda}$ is the Kronecker delta (equal to 1 if $\Lambda = 0$ and 0 otherwise), $L_{A,B}$ and $S_{A,B}$ denoting the orbital and spin angular momenta of the colliding atoms A and B, respectively, and $S'$ the spin multiplicity of the $\Lambda$ state. The matrix element $M_{EJ,v'J'}$ is given by \cite{johnson1977new}:
\begin{equation}\label{eq3}
M_{EJ,v'J'} = \int_0^\infty \phi_J(E,R) \, D(R) \, \psi_{v'J'}(R) \, dR,
\end{equation}
where $D(R)$ is the transition dipole moment function, $\phi_J(E, R)$ is the energy-normalized continuum radial wavefunction of the initial state, and $\psi_{v'J'}(R)$ is the bound radial wavefunction of the final state.

\begin{table}[htpb]
\caption{H\"{o}nl-London factors for selected transitions.}
\centering
\begin{tabular}{ccc}
\hline
   & ${^1\Sigma^+} \to {^1\Sigma^+}$ & ${^1\Pi} \to {^1\Sigma^+}$  \\
\hline
P-branch & $J$ & $(J+1)/4$   \\
Q-branch &   --    & $(2J+1)/4$ \\
R-branch & $J+1$ & $J/4$  \\
\hline
\end{tabular}
\label{tab:HL}
\end{table}

To account for stimulated emission induced by a blackbody radiation field at temperature $T_b$, the partial cross section for each rovibrational transition $(v',J')$ is enhanced by the stimulation factor \cite{stancilStim1997}:
\begin{equation}
\sigma_{\Lambda \to \Lambda'}(E;T_b) = \frac{\sigma_{\Lambda \to \Lambda'}^{\rm sp}(E)}{1 - \exp(-h\nu_{E,v',J'}/k_B T_b)}.
\end{equation}
The photon occupation number in the blackbody field provides the additional stimulated contribution, which becomes significant at higher radiation temperatures or for transitions with lower frequencies $\nu_{E,v',J'}$. The total cross section is obtained by summing the enhanced partial contributions over all accessible $(v',J')$ levels. When $T_b = 0$ K, the stimulation factor reduces to unity and the expression recovers the pure spontaneous case.

The thermal rate coefficients are obtained by averaging the cross sections over a Maxwell--Boltzmann velocity distribution:
\begin{equation}
k_{\Lambda \to \Lambda'}(T) = \sqrt{\frac{8}{\mu\pi(k_B T)^3}} \int_0^\infty E\,\sigma_{\Lambda \to \Lambda'}(E)\, e^{-E/k_B T}\, dE,
\end{equation}
where $T$ is the gas temperature, $\mu$ the reduced mass, and $k_B$ the Boltzmann constant.

\section{Results and Discussion}\label{S3}

\subsection{Electronic structure of AgH}\label{S3.1}

\begin{table*}[!htbp] 
\renewcommand{\arraystretch}{1}       
\small
\centering    
    \caption{Molecular states of AgH correlating with the lowest dissociation limits, together with the asymptotic energies, long-range fitting coefficients of the PECs and TDMs (PDM for X$^1\Sigma^+$), and the statistical weights $p_\Lambda$. All coefficients are in atomic units; numbers in parentheses denote powers of 10.}    \setlength{\tabcolsep}{3.65pt}
\begin{tabular}{l c c c c c c c c c c c}   
\hline \hline
\multirow{2}{*}{Asymptotic atomic states} &  \multirow{2}{*}{State} &  {NIST} & {Molpro} & \multicolumn{3}{c} { PEC fitting Coefficient}  & & \multicolumn{3}{c}{ TDM(PDM) fitting Coefficient} & \multirow{2}{*}{$p_{\Lambda}$} \\
\cline {5-7}\cline {9-11} &  & (eV) & (eV)  & A  & B & C & & D  & E & F   & \\
 \hline
$\mathrm{Ag}(4d^{10}5s~^2\mathrm{S})+\mathrm{H}(1s~^2\mathrm{S})$  &$\mathrm{X^1\Sigma ^+}$ & 0.000  &  0.000 & 7.617(3) & 8.767(1) & -9.687(4) & & 4.221(1) & -8.411(4) & 4.184(7) & 1/4 \\
$\mathrm{Ag}(4d^{10}5p~^2\mathrm{P}^{\circ})+\mathrm{H}(1s~^2\mathrm{S})$   & $\mathrm{A^1\Sigma ^+}$  & 3.664   &3.698& 2.689(2) & 1.832(2) & -2.224(5)  & & -3.442(2) & 5.632(5) & -1.836(8)  & 1/12\\
   & $\mathrm{1^1\Pi} $ & 3.664   &3.696& 1.784(1) & -2.176(2) & 2.893(4)  & & 1.175(1) & -1.965(4) & 6.895(6)  & 1/6\\
$\mathrm{Ag}(4d^95s^2~^2\mathrm{D})+\mathrm{H}(1s~^2\mathrm{S})$  &$\mathrm{3^1\Sigma ^+}$ & 3.750   & 3.795 & 1.486(1) & -2.002(2) & 3.623(4) &  &-1.590(1) & 2.509(4) & -7.466(6) & 1/20 \\
    & $\mathrm{2^1\Pi} $ & 3.750   & 3.795& -2.996(2) & 1.204(2) & -1.035(5)  &  & 1.540(2) & -2.473(5) & 7.574(7)   & 1/10\\
\hline  
\hline                  
\end{tabular}
\label{t1}  
\end{table*}

\begin{figure}[!htbp] 
    \centering
    \includegraphics[width=0.48\textwidth]{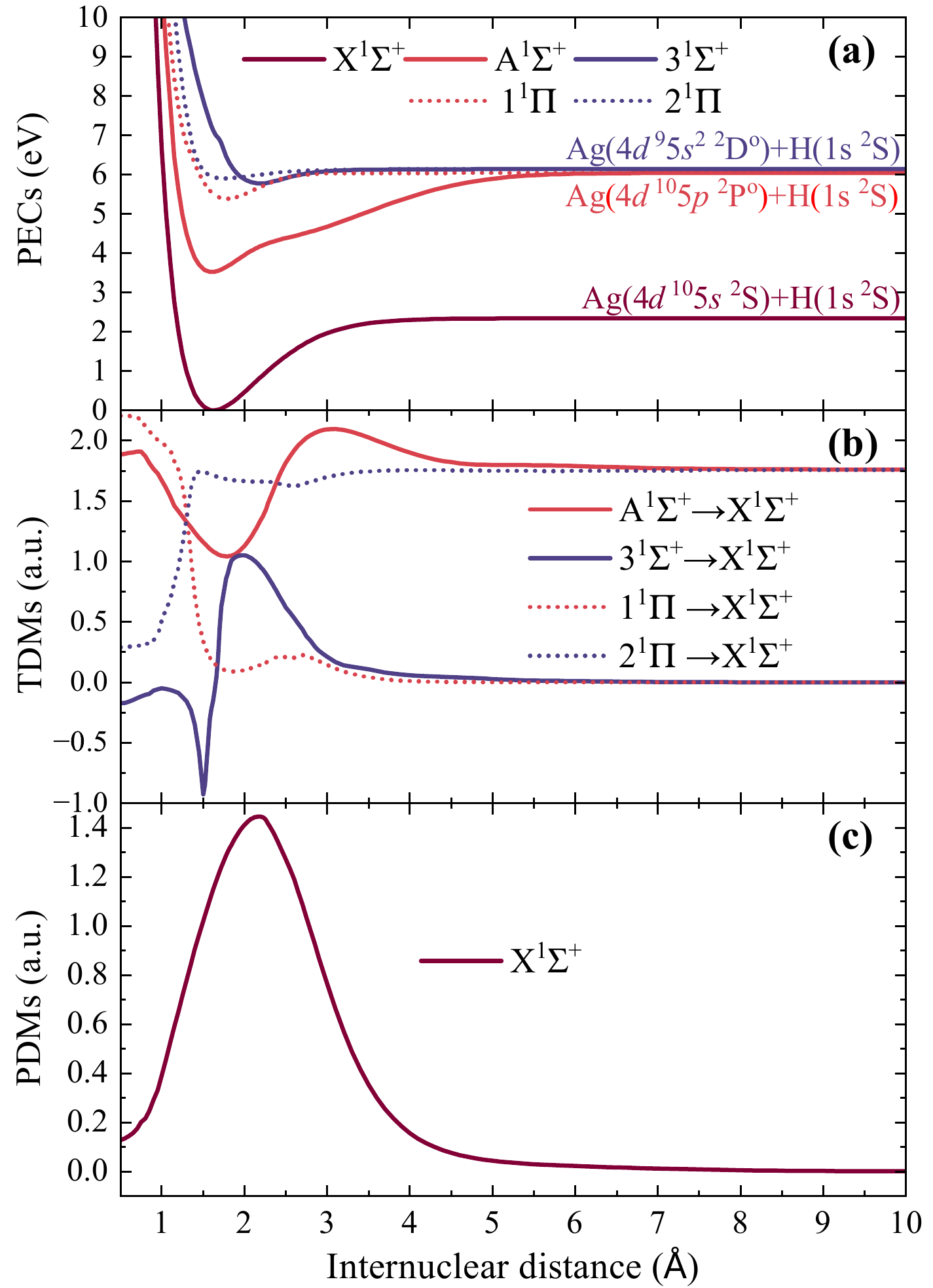} 
    \caption{Potential energy curves for the five low-lying $\Lambda-S$ states of AgH, transition dipole moments from excited states to X$^1\Sigma^+$, and permanent dipole moment of the X$^1\Sigma^+$ ground state.}
    \label{fig1}
\end{figure}

The calculated PECs of the five low-lying singlet states, together with the transition dipole moments (TDMs) from the excited states to X$^1\Sigma^+$ and the permanent dipole moment (PDM) of the X$^1\Sigma^+$ ground state, are displayed in Figure~\ref{fig1}. The corresponding long-range fitting coefficients of the PECs and dipole moments, as well as the statistical weights $p_\Lambda$ relevant to the radiative association channels, are summarized in Table~\ref{t1}. The calculated asymptotic energies agree well with the NIST values \cite{moore1949atomic,badr2004continuous,badr2006observation}, with deviations of 0.034~eV for the $\mathrm{Ag}(4d^{10}5p)+\mathrm{H}(1s)$ limit and 0.045~eV for the $\mathrm{Ag}(4d^95s^2)+\mathrm{H}(1s)$ limit, supporting the reliability of the present MRCI+Q treatment.

The X$^1\Sigma^+$ and A$^1\Sigma^+$ states exhibit relatively deep wells, whereas the $1^1\Pi$, $3^1\Sigma^+$, and $2^1\Pi$ states are shallower. This difference is important for the subsequent RA dynamics, since it directly affects the density and character of the quasi-bound levels in the entrance channels. In the dipole-moment curves, the A$^1\Sigma^+$--X$^1\Sigma^+$ and $3^1\Sigma^+$--X$^1\Sigma^+$ transitions show abrupt changes near 1.8~\AA, while the $1^1\Pi$--X$^1\Sigma^+$ and $2^1\Pi$--X$^1\Sigma^+$ transitions vary sharply near 1.3~\AA. These features are associated with weak avoided crossings in the corresponding excited-state PECs. The PDM of X$^1\Sigma^+$ increases first and then gradually approaches zero at large $R$, reflecting the transition from the bonded region to the neutral separated atom limit.

\begin{table}[h]
\centering
\caption{Spectroscopic constants of the low-lying electronic states of AgH.}
\renewcommand{\arraystretch}{1}
\setlength{\tabcolsep}{6pt}

\begin{tabular}{l c c c c c}
\hline
\multirow{2}{*}{State} & $R_e$ & $\omega_e$ & $\omega_e\chi_e$ & $B_e$ & $D_e$ \\
\multicolumn{1}{c}{} & (\AA) & (cm$^{-1}$) & (cm$^{-1}$) & (cm$^{-1}$) & (eV) \\
\hline
X$^1\Sigma^+$  & 1.62$^{e}$ & 1749.58$^{e}$ & 35.42$^{e}$ & 6.19$^{e}$ & 2.34$^{e}$ \\
              & 1.61$^{a}$ & 1767.47$^{a}$ & 36.38$^{a}$ & 6.48$^{a}$ & 2.36$^{a}$\\
              & 1.62$^{b}$ & 1759.90$^{b}$  & 34.18$^{b}$  & 6.50$^{b}$  & 2.38$^{b}$   \\
A$^1\Sigma^+$  & 1.60$^{e}$ & 1577.78$^{e}$ & 53.38$^{e}$ & 6.27$^{e}$ & 2.52$^{e}$  \\
              & 1.62$^{a}$ & 1812.19$^{a}$ & 63.20$^{a}$ & 6.38$^{a}$ & 2.36$^{a}$ \\
              & 1.64$^{c}$  & 1663.60$^{c}$  & 87.00$^{c}$     & 6.27$^{c}$ & 2.36$^{c}$  \\
$1^1\Pi$      & 1.78$^{e}$ & 1480.50$^{e}$ & 85.60$^{e}$ & 5.49$^{e}$ & 0.66$^{e}$  \\
              & 1.79$^{d}$ & 1477.00$^{d}$ & 62.00$^{d}$ & 5.28$^{d}$ & 0.63$^{d}$  \\
$2^1\Pi$      & 1.75$^{e}$ & 852.00$^{e}$ & 90.33$^{e}$ & 4.76$^{e}$ & 0.24$^{e}$  \\
$3^1\Sigma^+$  & 2.20$^{e}$ & 1264.00$^{e}$ & 130.77$^{e}$ & 3.78$^{e}$ & 0.37$^{e}$ \\
\hline
\end{tabular}
\label{tab3}

\vspace{5pt}
\parbox{\linewidth}{\small
\textbf{Note.}
$a$: \citet{zhou2023theoretical};
$b$: \citet{le2005direct};
$c$: \citet{learner1962influence};
$d$: \citet{li2018candidates};
$e$: present work.
}
\end{table}

To further assess the quality of the present electronic-structure results, Table~\ref{tab3} compares the calculated spectroscopic constants with available theoretical \cite{zhou2023theoretical,li2018candidates} and experimental \cite{le2005direct,learner1962influence}   data. For the X$^1\Sigma^+$ ground state, the present values of $R_e$, $\omega_e$, $\omega_e\chi_e$, $B_e$, and $D_e$ are in good agreement with previous high-level calculations and spectroscopic measurements \cite{zhou2023theoretical,le2005direct}. The same is true, at least qualitatively, for the excited singlet states, considering that some of them are relatively shallow and more difficult to characterize accurately.  The overall agreement is satisfactory, confirming that the present electronic structure results provide a reliable basis for the radiative association calculations discussed below.


\subsection{Vibrationally and rotationally resolved cross sections}\label{S3.2}

\begin{figure}[!htbp] 
\vspace{2.2mm}
    \centering
    \includegraphics[width=0.47\textwidth]{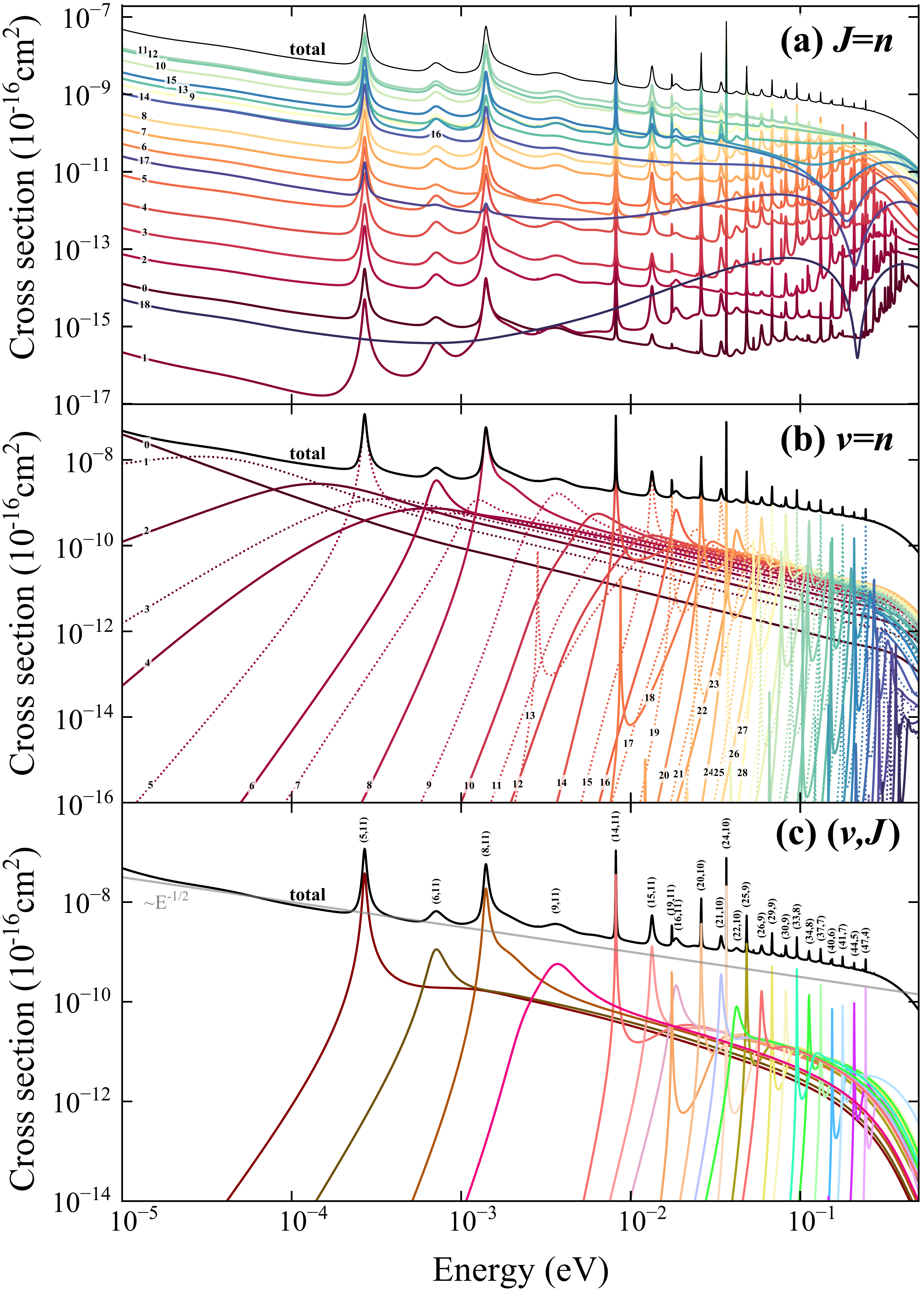} 
    \caption{Vibrationally and rotationally resolved radiative association cross sections at T = 0 K for the $\mathrm{X^1\Sigma^+ \to X^1\Sigma^+}$ channel in AgH. (a) Partial cross sections summed over vibrational quantum numbers $v$ for fixed rotational quantum numbers $J$ ($J=n$), compared to the total cross section; (b) Partial cross sections summed over $J$ for fixed $v$ ($v=n$); (c) Contributions from specific $(v, J)$ ro-vibrational states, with dominant resonances labeled.}
    \label{fig2}
\end{figure}

The radiative association (RA) cross sections for the Ag + H collision system were computed using a full quantum mechanical approach, incorporating the high-precision \emph{ab initio} PECs and TDMs/PDM. As an illustrative example, Figure~\ref{fig2} presents the vibrationally and rotationally resolved cross sections for the ground-state X$^1\Sigma^+ \to \text{X}^1\Sigma^+$ channel, corresponding to spontaneous RA ($T=0$ K). Panel (a) displays the partial cross sections summed over all vibrational quantum numbers $v$ for fixed rotational quantum numbers $J$ ($J=n$), compared to the total cross section; panel (b) shows the sums over $J$ for fixed $v$ ($v=n$); and panel (c) highlights contributions from specific $(v, J)$ ro-vibrational states. The cross sections exhibit sharp resonances, particularly in the $10^{-4}$--$5\times10^{-1}$ eV, arising from quasi-bound states formed behind centrifugal barriers in the effective potential.


\begin{figure*}[!htbp] 
    \centering
    \includegraphics[width=0.86\textwidth]{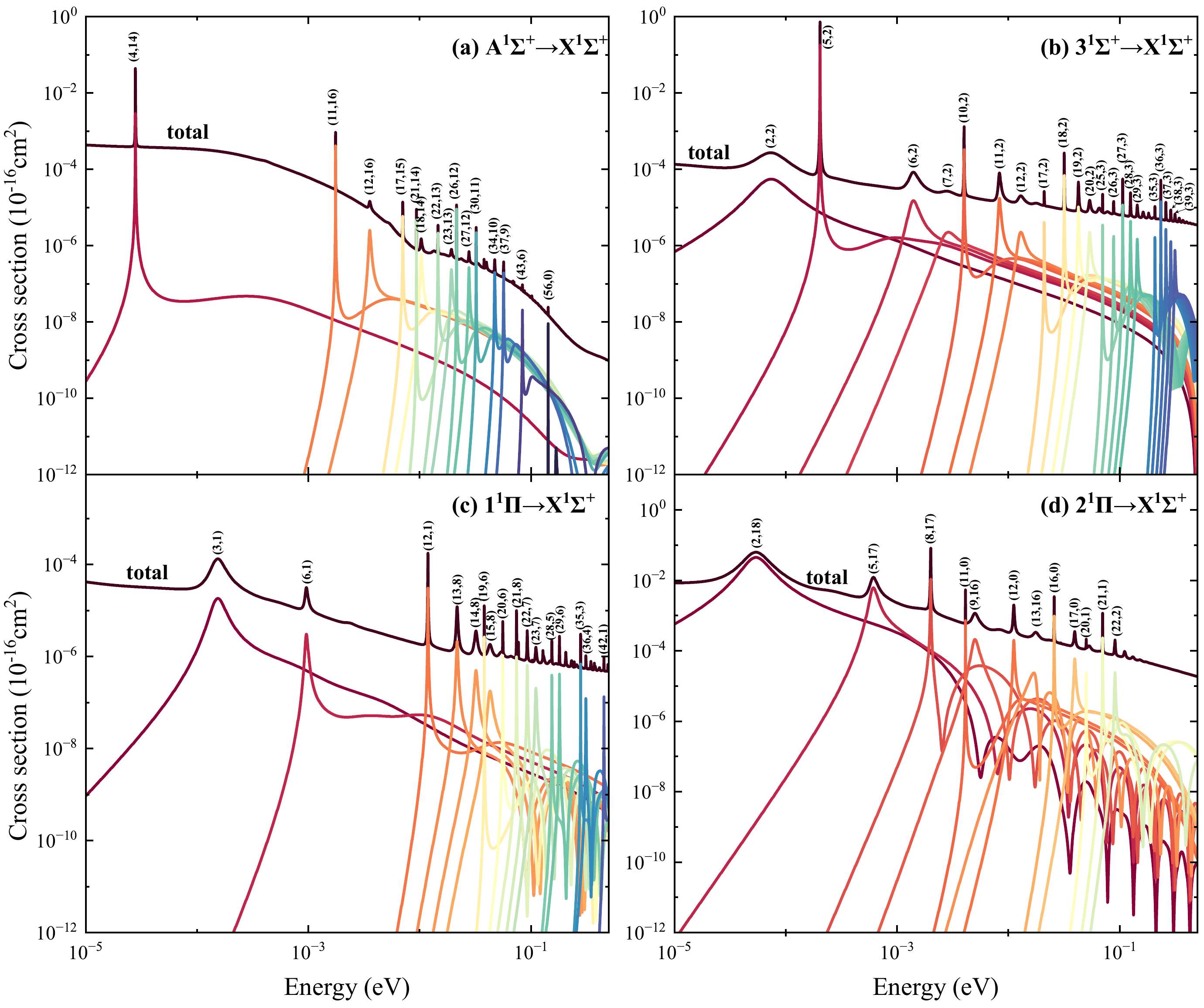} 
    \caption{Vibrationally and rotationally resolved radiative association cross sections for transitions from excited states to the ground state $\mathrm{X^1\Sigma^+}$ in AgH at $T=0$ K (spontaneous process).}

     \label{fig3}
\end{figure*}

Analysis of figure~\ref{fig2} panel (a) reveals that the dominant contributions to the total cross section stem from $J=9$--$16$, with $J=11$ providing the largest enhancement due to optimal centrifugal barrier effects that trap the colliding partners, promoting resonance formation. Each $J$ contributes resonances across the energy range, but as $J$ increases, the number of resonances diminishes, nearly vanishing by $J=18$, as the stronger centrifugal repulsion shifts the effective potential, reducing accessible quasi-bound levels at higher energies. 
Combining this with Figure~\ref{fig2} panel (b), which shows sums over $J$ for fixed $v$, indicates that these resonances are primarily driven by individual vibrational modes $v$. Notably, the resonance peaks at specific energy positions are dominated by higher $v$ values, exhibiting a progressive relationship where peak positions shift to higher energies with increasing $v$. In contrast, the low-energy region (less than $10^{-4}$ eV) is primarily contributed by $v=0$.
Figure~\ref{fig2} panel (c) confirms that specific $(v, J)$ states, such as $(5,11)$ and $(6,11)$, account for the major resonances. Notably, as the energy increases, the dominant contributions shift from $J=11$ to progressively lower values such as $J=10$, $9$, $8$, and $7$, indicating that lower angular momentum channels become more influential at higher energies; conversely, $v$ gradually increases from 5 to 47, reflecting the accessibility of higher vibrational states at elevated energies. This is consistent with the observations in panels (a) and (b). Physically, this reflects the reduced centrifugal barrier for low-$J$ states at elevated collision energies, allowing easier access to quasi-bound levels and radiative stabilization compared to high-$J$ channels, where the stronger repulsive centrifugal potential suppresses contributions. Furthermore, non-resonant background contributions follow a smooth $E^{-1/2}$ decay, while resonances amplify $\sigma$ by up to one or two orders of magnitude. These shape resonances enhance the RA efficiency in astrophysical low-temperature environments \cite{zygelman1998}.


Building on the ground-state analysis, the RA cross sections from the excited states (A$^1\Sigma^+$, $1^1\Pi$, $3^1\Sigma^+$, and $2^1\Pi$) to the ground state at $T=0$ K (Figure~\ref{fig3}) show distinct magnitudes and energy dependences, while all exhibit resonance enhancement. The underlying mechanism is similar to that in the ground-state continuum-to-bound process. The main difference arises from the higher dissociation limits of the excited states, which shift the resonance thresholds and modify the Franck--Condon factors. 
Due to the differing PEC well depths, shallower wells in states like $1^1\Pi$, $3^1\Sigma^+$, and $2^1\Pi$ promote stronger overlap with the incoming continuum wave functions and broader resonances. For instance, the first peak in panel (d) for $\mathrm{2^1\Pi \to X^1\Sigma^+}$ at $(2,8)$, while the deeper well in $\mathrm{A^1\Sigma^+}$ yields more localized quasi-bound states, resulting in narrower features, as seen in the first peak in panel (a) at $(4,14)$. As energy increases, the resonance peaks gradually narrow, reflecting a faster decay from weak quasi-bound states at high kinetic energies. 
Each panel shows the total cross section together with the dominant vibrationally and rotationally resolved contributions, labeled by specific $(v,J)$ pairs. As in the ground-state channel, these resonances arise from quasi-bound rovibrational levels supported by the initial-state PECs.
Their positions and intensities depend on the electronic-state character and on the associated rovibrational quantum numbers.
As collision energy rises, the cross sections from individual ro-vibrational channels gradually diminish, leading to an overall smooth decay in the total RA cross section. These results highlight the pivotal role of the initial electronic state and its ro-vibrational structure in the RA dynamics of AgH, where different electronic states significantly promote molecular formation through quasi-bound ro-vibrational state generation.


\subsection{Total cross sections and stimulated radiative association}\label{S3.3}

\begin{figure}[!htbp]
    \centering
    \includegraphics[width=0.48\textwidth]{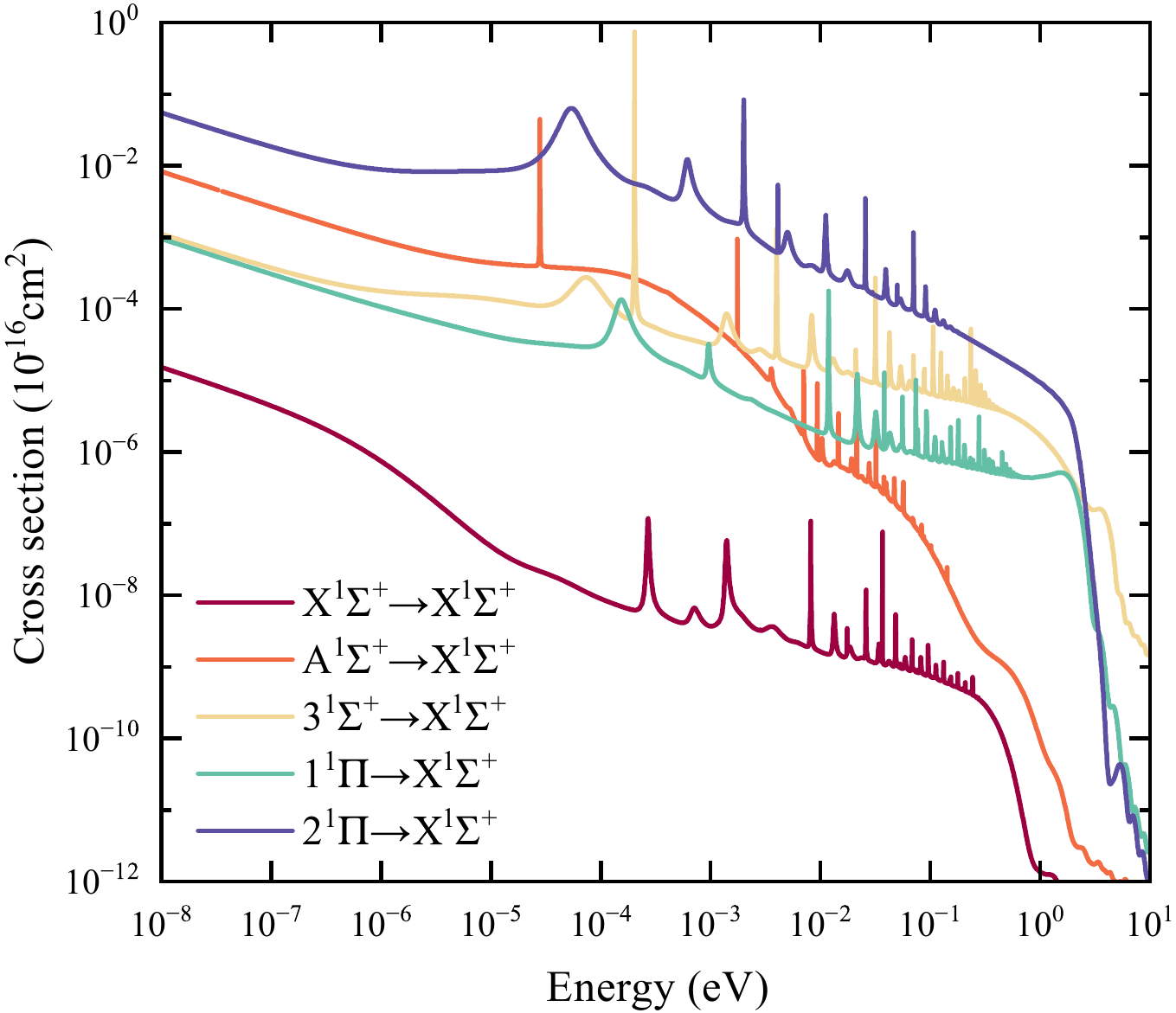} 
    \caption{Radiative association cross sections for transitions from the initial states X$^1\Sigma^+$, A$^1\Sigma^+$, $3^1\Sigma^+$, $1^1\Pi$, and $2^1\Pi$ to the ground state X$^1\Sigma^+$ in AgH, calculated at $T=0$ K.}
      \label{fig4}
\end{figure}

Figure~\ref{fig4} illustrates the total RA cross sections at T = 0 K for AgH formation via radiative transitions from various initial electronic states to the ground state X$^1\Sigma^+$ , as a function of collision energy. Notably, the $\mathrm{2^1\Pi \to X^1\Sigma^+}$ channel dominates the overall profile, owing to its favorable TDM and PEC characteristics that enhance continuum-bound overlaps. Resonances manifest in diverse shapes across the channels, reflecting the presence of quasi-bound states supported by the initial electronic effective potentials; these states significantly extend the interaction time of the colliding system, thereby substantially boosting the RA probability. As collision energy increases to around 1 eV, all channels exhibit a rapid decay in RA cross sections, indicating that under high-energy conditions, the system struggles to effectively dissipate excess kinetic energy via radiative transitions, thus hindering stable molecule formation. Overall, the energy-dependent behavior of AgH's RA cross sections directly embodies the modulating role of the initial electronic PEC structures on quasi-bound states and resonance effects, providing critical insights into molecular formation kinetics in astrophysical contexts.


\begin{figure}
    \centering
    \includegraphics[width=0.48\textwidth]{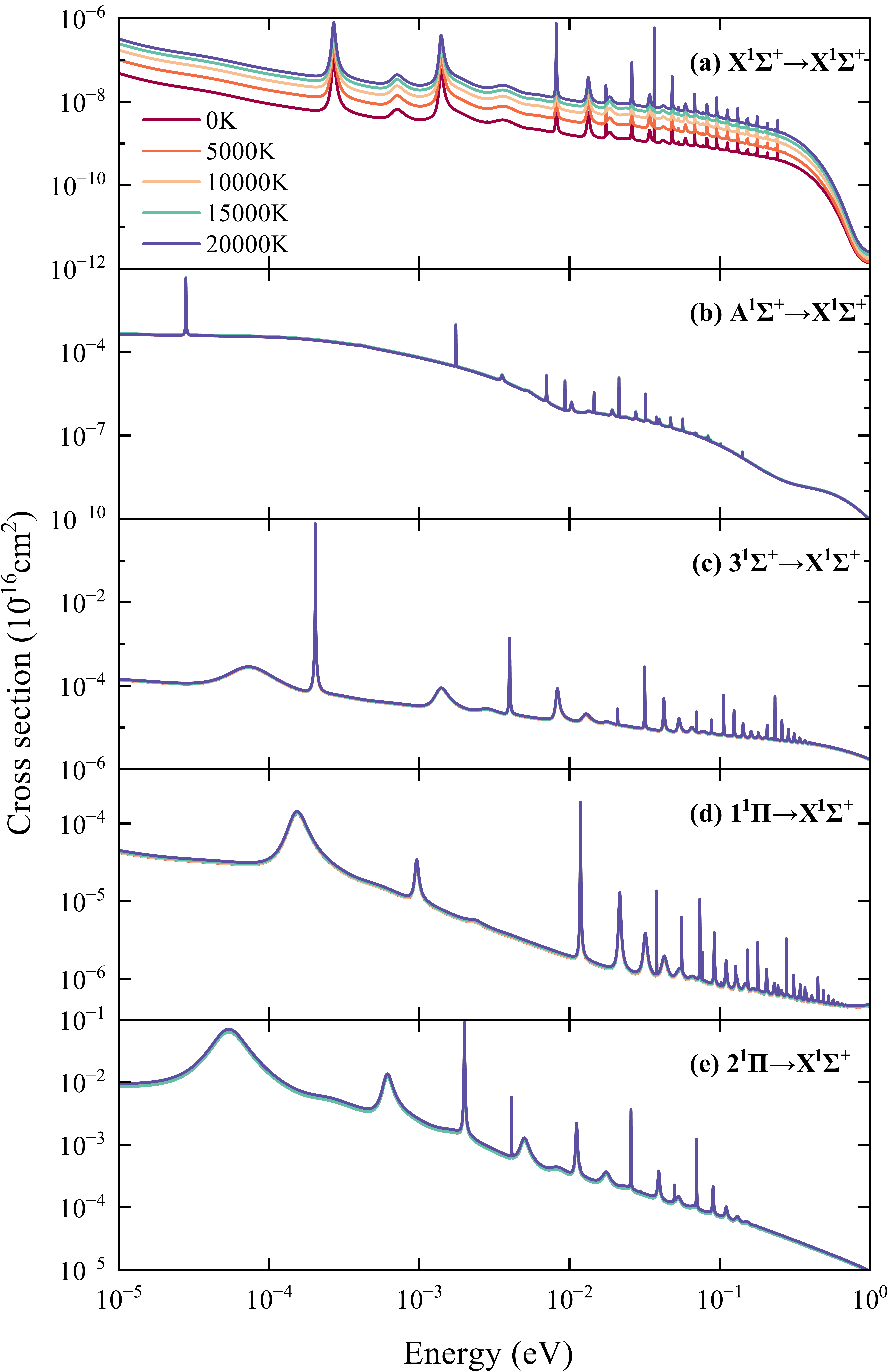} 
    \caption{Stimulated radiative association cross sections for AgH formation via transitions from the initial electronic states X$^1\Sigma^+$, A$^1\Sigma^+$, $3^1\Sigma^+$, $1^1\Pi$, and $2^1\Pi$ to the ground state X$^1\Sigma^+$ under blackbody radiation fields with temperatures of 0, 5000, 10000, 15000, and 20000 K.}
     \label{fig5}
\end{figure}

To assess the influence of blackbody radiation fields, we computed stimulated RA cross sections for AgH formation from various initial electronic states to the ground state X$^1\Sigma^+$, with radiation temperatures up to 20,000 K. As shown in Figure~\ref{fig5}, the overall energy dependence and resonance structures remain essentially unchanged relative to the spontaneous case ($T=0$ K), indicating that the RA dynamics are primarily governed by the underlying PECs and TDMs. 

For the excited-state channels (panels b--e), the cross sections are nearly insensitive to the radiation temperature over the range considered. In contrast, the ground-state channel X$^1\Sigma^+ \to$ X$^1\Sigma^+$ (panel a) shows a systematic enhancement in magnitude while retaining the same energy profile. The ratios of the stimulated to spontaneous cross sections are approximately 2.12, 3.62, 5.15, and 6.69 at radiation temperatures of 5000, 10000, 15000, and 20000 K, respectively. Thus, blackbody radiation primarily rescales the RA cross sections without altering their intrinsic resonance structures, and its effect is significant only for the ground-state channel.


\subsection{Thermal rate coefficients}\label{S3.4}

\begin{figure}
    \centering
    \includegraphics[width=0.48\textwidth]{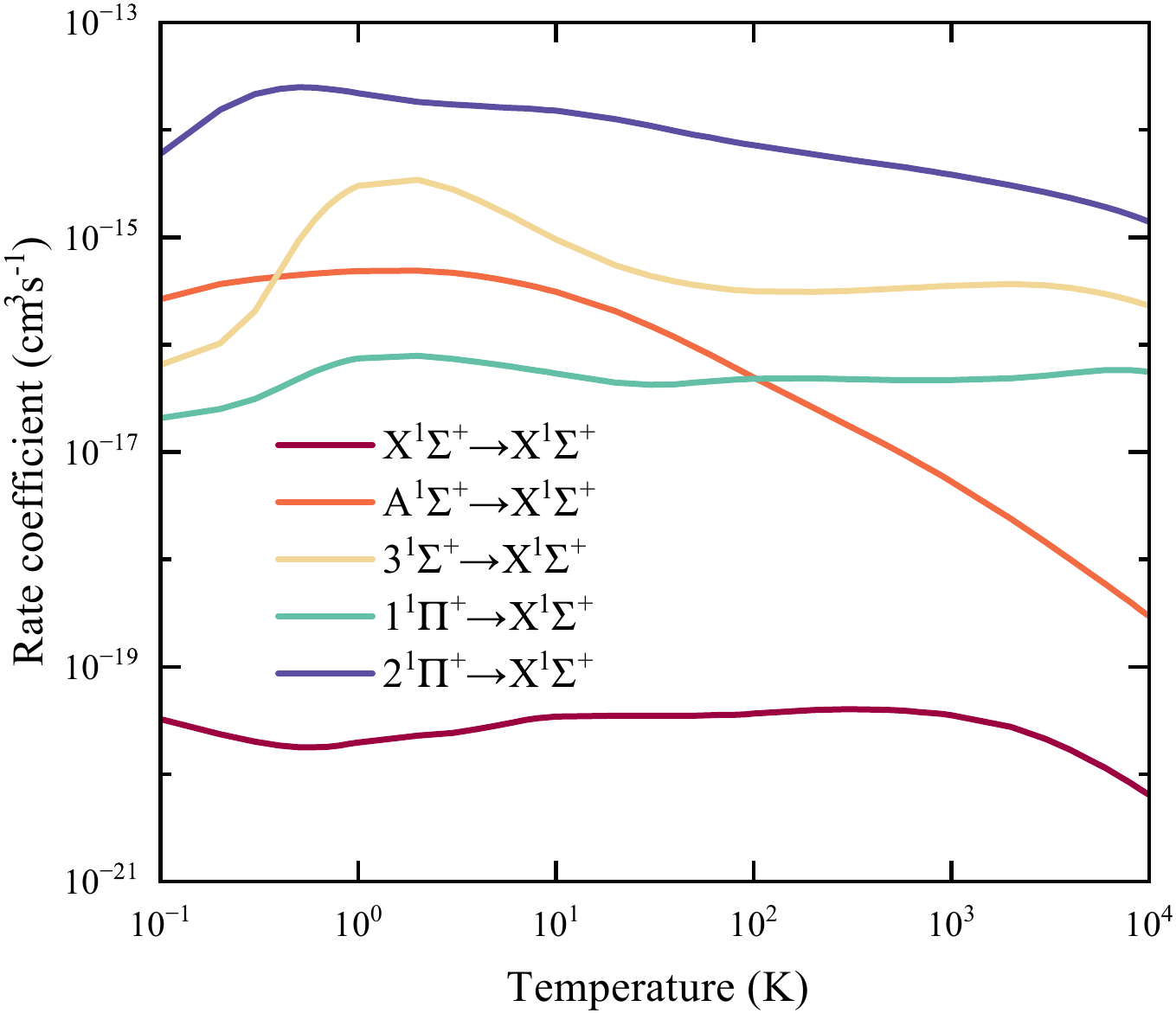} 
    \caption{Spontaneous radiative association (T = 0 K) rate coefficients for AgH formation via transitions from various initial electronic states to the ground state $\mathrm{X^1\Sigma^+}$, as a function of temperature.}
     \label{fig6}
\end{figure}

The thermal rate coefficients for AgH formation via radiative association were obtained by integrating the cross sections over a Maxwell--Boltzmann distribution in the temperature range $10^{-1}$--$10^4$~K, as shown in Figure~\ref{fig6}. The $\mathrm{2^1\Pi \to X^1\Sigma^+}$ channel yields the largest rate coefficient, reaching $\sim10^{-14}$~cm$^3$\,s$^{-1}$ at low temperatures, consistent with its dominant cross section. All channels show a general decrease with increasing temperature, reflecting the reduced efficiency of radiative stabilization at higher collision energies.

At low temperatures ($T \le 10$~K), the rate coefficients are dominated by resonance-rich low-energy cross sections. The A$^1\Sigma^+$ and $3^1\Sigma^+$ channels have comparable magnitudes ($\sim10^{-16}$--$10^{-15}$~cm$^3$\,s$^{-1}$), followed by the $1^1\Pi$ channel, whereas the X$^1\Sigma^+ \to$ X$^1\Sigma^+$ channel remains the weakest. Since the stimulated RA cross sections preserve the same energy dependence as in the spontaneous case (Figure~\ref{fig5}), the corresponding thermal rate coefficients at finite radiation temperatures can be obtained by applying multiplicative scaling factors to the $T=0$ K results. For the excited-state channels, these factors remain close to unity and can be neglected. For the ground-state channel, the rate coefficients are enhanced by factors of 2.12, 3.62, 5.15, and 6.69 at radiation temperatures of 5000, 10000, 15000, and 20000 K, respectively.

At temperatures relevant to circumstellar envelopes and diffuse clouds ($\sim$10--1000~K), the dominant rate coefficients are of the order of $10^{-16}$--$10^{-14}$~cm$^3$\,s$^{-1}$, comparable to those reported for analogous species such as AlO \cite{Di2025} and MgS \cite{Wei2025}, indicating that radiative association is a viable pathway for AgH formation in low-density astrophysical environments. 
The complete dataset of cross sections and rate coefficients is publicly available at Zenodo (doi:10.5281/zenodo.19041836) for use in astrochemical modeling.

\section{Conclusions}\label{S4}

Radiative association processes in the Ag + H collision system have been investigated in detail for the first time using full quantum scattering calculations supported by high-accuracy \emph{ab initio} potential energy curves and transition dipole moments for the low-lying electronic states X$^1\Sigma^+$, A$^1\Sigma^+$, $1^1\Pi$, $3^1\Sigma^+$, and $2^1\Pi$.
The computed vibrationally and rotationally resolved cross sections reveal prominent shape resonances at ultralow collision energies ($10^{-4}$--$1$ eV), originating from quasi-bound rovibrational levels trapped behind centrifugal barriers in the entrance channels. These resonances significantly enhance association probabilities, with contributions from intermediate rotational quantum numbers ($J \approx 9$--$16$) proving particularly effective in the ground-state channel due to optimal barrier heights that prolong interaction times.
Among the examined transitions to the ground X$^1\Sigma^+$ state, the $2^1\Pi \to$ X$^1\Sigma^+$ channel emerges as the dominant pathway across most of the energy range. Shallower potential wells in states such as $1^1\Pi$, $3^1\Sigma^+$, and $2^1\Pi$ yield broader resonances compared to the deeper A$^1\Sigma^+$ well. 
Stimulated radiative association under blackbody radiation fields (up to $20\,000$ K) primarily rescales the cross sections without modifying the intrinsic resonance structures. This effect is significant only for the ground-state X$^1\Sigma^+ \to$ X$^1\Sigma^+$ channel, whereas the excited-state channels remain nearly unchanged.
The thermal rate coefficients, computed over 10$^{-1}$--$10^4$~K, show a general decreasing trend with increasing temperature for all channels, reflecting the decline of RA cross sections at higher collision energies. The dominant $2^1\Pi \to$ X$^1\Sigma^+$ channel reaches $\sim10^{-14}$~cm$^3$\,s$^{-1}$ at low temperatures, comparable in magnitude to radiative association rates reported for other astrophysically relevant diatomic species such as AlO and MgS.
These results address a critical gap in radiative association data for transition-metal hydrides, providing reliable rate coefficients essential for astrochemical modelling of AgH formation in low-temperature environments, including diffuse interstellar clouds, circumstellar outflows from AGB stars, and cool stellar atmospheres.

\section*{Acknowledgments}

This work was supported by the National Natural Science Foundation of China (Grant Nos. 12504297, 12304279, 12274040, 62475182), 
the National Key R\&D Program of China (Grant No. 2022YFA1602504), 
the special fund for Science and Technology Innovation Teams of Shanxi Province (202304051001034), 
the Special project for science and technology cooperation and exchange of Shanxi Province (202304041101022),
the Key Research and Development Program of Shanxi Province of China (202402150301012, 202402130501005, 202302150101006),
the Shanxi Province Scientific Research Initial Funding (Grant No. 20252011) and 
the Taiyuan University of Science and Technology Scientific Research Initial Funding (Grant No. 20242149).

\bibliographystyle{unsrt}  
\bibliography{references}

\end{document}